\documentclass[11pt]{article}

\usepackage{amsmath,amssymb,color}
\usepackage{pstcol}
\usepackage{graphicx}

\newcommand{\p}{\ensuremath{\partial}}

\newcommand\shalf{\ensuremath{{\scriptstyle\frac{1}{2}}}}

\newcommand{\rem}[1]{}
\newtheorem{theorem}{Theorem}
\newtheorem{lemma}{Lemma}

\DeclareMathAlphabet{\mathbi}{OML}{cmm}{b}{it} 
\newcommand{\bx}{\mathbi{x}}

\newcommand{\bel}{\begin{equation}\label}
\newcommand{\ee}{\end{equation}}
\newcommand{\ben}{\begin{enumerate}}
\newcommand{\een}{\end{enumerate}}
\newcommand{\bde}{\begin{description}}
\newcommand{\ede}{\end{description}}
\newcommand{\bit}{\begin{itemize}}
\newcommand{\eit}{\end{itemize}}
\newcommand{\bc}{\begin{center}}
\newcommand{\ec}{\end{center}}
\newcommand{\bB}{\mbox{\boldmath$B$}}
\newcommand{\bhB}{\mbox{\boldmath$\hat{B}$}}
\newcommand{\bdb}{\mbox{\boldmath$b$}}
\newcommand{\bD}{\mbox{\boldmath$\mathcal{D}$}}
\newcommand{\bh}{\mathbi{\hat{b}}}
\newcommand{\ba}{\mathbi{a}}

\newcommand{\br}{\mathbi{r}}
\newcommand{\bR}{\mathbi{R}}
\newcommand{\bfR}{\mathfrak{R}}
\newcommand{\bs}{\mathbi{s}}
\newcommand{\bq}{\mathbi{q}}
\newcommand{\bqr}{\mbox{\boldmath$\omega_{\rho}$}}
\newcommand{\bqt}{\mbox{\boldmath$\tilde{\omega}$}}

\newcommand{\bfq}{\mathfrak{q}}
\newcommand{\bhfq}{\hat{\mathfrak{q}}}
\newcommand{\bhfp}{\hat{\mathfrak{p}}}

\newcommand{\bfw}{\mathfrak{w}}

\newcommand{\bfr}{\mathfrak{r}}

\newcommand{\cast}{\circledast}

\newcommand{\bmu}{\mbox{\boldmath$\mu$}}

\newcommand{\bdl}{\mbox{\boldmath$\delta\ell$}}
\newcommand{\bhn}{\hat{\mathbi{n}}}
\newcommand{\bu}{\mathbi{u}}
\newcommand{\bv}{\mathbi{v}}

\newcommand{\bQ}{\mathbi{Q}}
\newcommand{\bw}{\mathbi{w}}
\newcommand{\bhw}{\mathbi{\hat{w}}}

\newcommand{\bPi}{\mathfrak{P}}
\newcommand{\bom}{\mbox{\boldmath$\omega$}}
\newcommand{\bhom}{\mbox{\boldmath$\hat{\omega}$}}
\newcommand{\bcapom}{\mbox{\boldmath$\Omega$}}

\newcommand{\bchi}{\mbox{\boldmath$\chi$}}
\newcommand{\bhchi}{\boldsymbol{\hat{\chi}}}
\newcommand{\bsigma}{\mbox{\boldmath$\sigma$}}

\newcommand{\beq}{\begin{eqnarray}} 
\newcommand{\eeq}{\end{eqnarray}} 
\newcommand{\nh}{\mathbi{\hat{n}}}
\makeatletter
\@addtoreset{equation}{section}

\makeatother

\begin{document}

\bc
\textbf{\Large Lagrangian particle paths \& ortho-normal quaternion frames}
\par\vspace{4mm}
\textbf{\large J. D. Gibbon$^1$ and D. D. Holm$^{1,2}$}
\par\vspace{.5cm}
$^1$Department of Mathematics, Imperial College London SW7 2AZ, UK
\\{\small email: j.d.gibbon@ic.ac.uk and d.holm@ic.ac.uk}
\par\vspace{3mm}
$^2$Computer and Computational Science,\\
Los Alamos National Laboratory,\\
MS D413 Los Alamos, NM 87545, USA\\
{\small email: dholm@lanl.gov}
\ec

\vspace{5mm}

\begin{abstract}
Experimentalists now measure intense rotations of Lagrangian particles in turbulent flows 
by tracking their trajectories and Lagrangian-average velocity gradients at high Reynolds 
numbers. This paper formulates the dynamics of an orthonormal frame attached to each 
Lagrangian fluid particle undergoing three-axis rotations, by using quaternions in 
combination with Ertel's theorem for frozen-in vorticity. The method is applicable to a 
wide range of Lagrangian flows including the three-dimensional Euler equations and its 
variants such as ideal MHD. The applicability of the quaterionic frame description to 
Lagrangian averaged velocity gradient dynamics is also demonstrated.
\end{abstract}

\vspace{1cm}

\tableofcontents

\newpage

\section{\large Introduction}\label{intro}

\vspace{-2mm}

\subsection{General background}\label{generalintro}

On 16th October, 1843, Hamilton wrote the expression for quaternion multiplication,
\[
\mathbb{I}^2=\mathbb{J}^2=\mathbb{K}^2
=\mathbb{I}\mathbb{J}\mathbb{K}
=-{\rm Id}
\,,
\]
as an algebraic rule for composing ``turns'' (directed arc lengths of great circles) in orienting  his telescope. 
This feature -- that multiplication of quaternions represents composition of 
rotations -- has made them the technical foundation of modern inertial guidance systems in 
the aerospace industry where tracking the paths and the orientation of satellites and aircraft 
is of paramount importance \cite{Ku99}. The graphics community also uses them to control the 
orientation of tumbling objects in computer animations because they avoid the difficulties 
incurred at the north and south poles when Euler angles are used \cite{Ha06}. 
\par\smallskip
Given the utility of quaternions in explaining the dynamics of rotating objects in flight, 
one might ask whether they would also be useful in tracking the orientation and angular 
velocity of Lagrangian particles in fluid dynamical situations. Recent experiments in 
turbulent flows have developed 
to the stage where the trajectories of tracer particles can be detected at high Reynolds 
numbers \cite{La01,M01,Vo02,MCB04,MLP04,M05,LTK05,Bif05,RMCB05,Eck06}\,; see Figure 1 in 
\cite{La01}. Numerical differentiation of these trajectories gives information about 
the Lagrangian velocity and acceleration of the particles. In particular the curvature 
of the particle paths can be used to extract statistical information about velocity 
gradients from a single trajectory \cite{Eck06}.
\par\smallskip
The usual practice in graphics problems is to consider the Frenet-frame of a trajectory 
which consists of the unit tangent vector, a normal and a bi-normal \cite{Ha06,Eck06}. 
In navigational language, this represents the corkscrew-like pitch, yaw and roll of 
the motion. In turn, the tangent vector and normals are related to the curvature and 
torsion. While the Frenet-frame describes the path, it ignores the dynamics that 
generates the motion. Here we will discuss another ortho-normal frame associated with 
the motion of each Lagrangian fluid particle, designated the \textit{quaternion-frame}. 
This may be envisioned as moving with the Lagrangian particles, but their evolution 
derives from the Eulerian equations of motion. 
\par\smallskip
The first main contribution of this paper lies in the explicit calculation of the 
evolution equations of this quaternion frame. Secondly, this formulation is shown 
to apply to a wide range of Lagrangian problems, as the next sub-section shows. Thirdly, 
the evolution of the quaternion frame is related to an associated Frenet frame. Finally,  
it is shown how the pressure Hessian, which plays a key role in driving the quaternion 
frame for Euler fluid flow, can be modelled using constitutive relations. 

\subsection{An appropriate class of Lagrangian flows}\label{lagclass}

Suppose $\bw$ is a contravariant vector quantity attached to a particle following a flow 
along characteristic paths $d\bx/dt=\bu(\bx,\,t)$ of a velocity $\bu$. Let us consider 
the abstract Lagrangian flow equation
\bel{w-dyn}
\frac{D\bw}{Dt} = \ba(\bx,\,t)\,,
\hspace{3cm}
\frac{D~}{Dt} = \frac{\p~}{\p t} + \bu\cdot\nabla\,.
\ee
where the material derivative has its standard definition. Examples of systems that 
(\ref{w-dyn}) might represent are\,:
\ben\itemsep -1mm
\item The vector $\bw$ in (\ref{w-dyn}) could be the velocity of a tracer particle 
in a fluid transported by a background velocity field $\bu$ with $\ba$ as the 
particle's acceleration. 

\item $\bw$ could be the vorticity $\bom = \hbox{curl}\,\bu$ of the incompressible 
Euler fluid equations, in which case $\ba = \bom\cdot\nabla\bu$ and $\hbox{div}\,\bu = 0$. 
The case of the Euler equations with rotation $\bcapom$ would make $\bw\equiv\bqt 
= \rho^{-1}(\bom + 2\bcapom)$.

\item For the barotropic compressible Euler fluid equations (where the pressure $p = p(\rho)$ 
is only density dependent) then $\bw \equiv \bqr = \rho^{-1}\bom$, in which case 
$\ba = \bqr\cdot\nabla\bu$ and $\hbox{div}\,\bu \ne 0$. 

\item The vector $\bw$ could represent a small vectorial line element $\bdl$ transported 
by a background flow $\bu$, in which case $\ba = \bdl\cdot\nabla\bu$.  For example, using 
Moffatt's analogy between the (magnetic) $\bB$-field in ideal MHD and fluid vorticity 
\cite{HKM1}, taking $\bdl \equiv \bB$ in the equations for incompressible ideal MHD, then 
$\ba = \bB\cdot\nabla\bu$ and $\hbox{div}\,\bB = 0$. In a slightly generalized form it 
could also represent the Elsasser variables $\bw^{\pm} = \bu \pm \bB$ in which case 
$\ba^{\pm} = \bw^{\pm}\cdot\nabla\bu$ with two material derivatives \cite{HKM1}. In 
each of the cases (2-4) the vectors $\bw$ and $\bu$ satisfy the standard Eulerian form
\bel{wev1}
\frac{D\bw}{Dt} = \bw\cdot\nabla\bu \,.
\ee
Consequently, it follows from Ertel's Theorem \cite{Er42,Ohk93,GHKR} that 
\bel{ertel2}
\frac{D(\bw\cdot\nabla\bmu)}{Dt} = \bw\cdot\nabla\left(\frac{D\bmu}{Dt}\right)\,,
\ee
for any differentiable function $\bmu(\bx,\,t)$. Choosing $\bmu = \bu(\bx,\,t)$ as in \cite{Ohk93} 
and identifying 
the flow acceleration as $D\bu/Dt =\bQ(\bx,\,t)$ yields the second derivative relation 
\bel{ertel3}
\frac{D^2\bw}{Dt^2} = \bw\cdot\nabla\left(\frac{D\bu}{Dt}\right) 
=: \bw\cdot\nabla\bQ\,.
\ee
\een
In each of the cases (2-4) above the acceleration vector $\bQ$ in (\ref{ertel3})  is readily identifiable. Thus, in those cases we have 
\bel{a-dyn}
\frac{D\bw}{Dt} = \ba(\bx,\,t)
\quad\hbox{and}\quad
\frac{D\ba}{Dt} = \bw\cdot\nabla\bQ =: \bdb(\bx,\,t)
\quad\hbox{along}\quad
\frac{D\bx}{Dt} = \bu(\bx,\,t)
\,.
\ee
These are the kinematic rates of change of the vectors $D\bw/Dt=\ba$ and $D\ba/Dt=\bdb$ 
following the characteristics of the velocity vector $\bu$ along the path $\bx(t)$ 
determined from $D\bx/Dt=\bu(\bx,\,t)$. 

\rem{
An alternative way of 
expressing Ertel's result is to say that since $\bw$ is contravariant, its components evolve 
under the coordinate transformation $\bx(0)\to\bx(t)$ along the characteristic path by the 
change of coordinates rule
\bel{contra-w}
\bw(t)\cdot\frac{\partial~~~~}{\partial\bx(t)}
= \bw(0)\cdot\frac{\partial~~~~}{\partial\bx(0)}\,.
\ee 
That is, the vector field $\bw\cdot\nabla$ is preserved along the characteristics
of the flow of the velocity vector $\bu$ (which is also contravariant). 
}
 
\par\smallskip
The plan of the paper is as follows:  \S\ref{lagev} shows that the quartet of 3-vectors 
$(\bu,\,\bw,\,\ba,\,\bdb)$ appearing in (\ref{a-dyn})
determines the quaternion-frame and its Lagrangian dynamics\,: indeed, a knowledge of $\bdb$ is 
essential to determining the dynamical process. Modulo a rotation around $\bw$, the quaternion-frame
turns out to be the Frenet-frame attached to lines  of constant $\bw$.  In case (2) where $\bw = \bom$,
lines  of constant $\bw$ are vortex lines and the particles are fluid parcels. As described in 
\cite{Ohk93} -- see also \cite{GHKR} for a history -- in this case Ertel's Theorem for Euler's 
fluid equations \cite{Er42} ensures that $\bdb = -{\sf P}\bom$ where ${\sf P}$ is the Hessian 
matrix of spatial derivatives of the pressure. In some practical situations, however, the vector $\bdb$ may not be known, or may not 
exist for every system for every triad $(\bu,\,\bw,\,\ba)$. For example no explicit relation for $\bdb$ is known for 
the Euler equations in velocity form for which $(\bu,\,\bw,\,\ba)\equiv (\bu,\,\bu,\,-\nabla p)$. 
\par\smallskip
Section \S\ref{examples} elaborates three examples: the Euler equations with rotation; the barotropic compressible Euler equations; and ideal MHD in Elsasser variables. 
Section \S\ref{constit} demonstrates the applicability of the quaterionic frame description to turbulence models of Lagrangian averaged velocity gradient dynamics, in which approximate constitutive relations for the pressure Hessian and auxiliary equations for the constitutive parameters are introduced. 

\section{\large Quaternions and rigid body dynamics}\label{rigid}

Rotations in rigid body mechanics has given rise to a rich and long-standing literature in 
which Whittaker's book is a classic example \cite{Whitt44}. This gives explicit formulae 
relating the Euler angles and what are called the Cayley-Klein parameters of a rotation. 
In fact the use of quaternions in this area is not only much more efficient but avoids 
the immensely complicated inter-relations that are unavoidable when Euler angle formulae 
are involved \cite{Klein04,Ta1890}. 
\par\smallskip
In terms of any scalar $p$ and any 3-vector $\bq$, the quaternion $\bfq = [p,\,\bq]$ is 
defined as (Gothic fonts denote quaternions)
\bel{quatdef1}
\bfq = [p,\,\bq] = p{\sf I} - \sum _{i=1}^{3}q_{i}\sigma_{i}\,,
\ee
where $\{\sigma_{1},\,\sigma_{2},\,\sigma_{3}\}$ are the three Pauli spin-matrices defined by
\bel{psm1}
\sigma_{1} = \left(\begin{array}{rr}
0 & i\\
i & 0
\end{array}\right)\,,
\hspace{1cm}
\sigma_{2} = \left(\begin{array}{rr}
0 & 1\\
-1 & 0
\end{array}\right)\,,
\hspace{1cm}
\sigma_{3} = \left(\begin{array}{rr}
i & 0\\
0 & -i
\end{array}\right)\,,
\ee
and ${\sf I}$ is the $2\times 2$ unit matrix. The relations between the Pauli matrices
$\sigma_{i}\sigma_{j} =  -\delta_{ij}{\sf I}-\epsilon_{ijk}\sigma_{k}$ then give a 
non-commutative multiplication rule
\bel{quatdef2}
\bfq_{1}\cast\bfq_{2} = [p_{1}p_{2} - \bq_{1}\cdot\bq_{2},\,p_{1}\bq_{2} + p_{2}\bq_{1} + 
\bq_{1}\times\bq_{2}]\,.
\ee
It can easily be demonstrated that quaternions are associative. 
\par\smallskip
Let $\bhfp = [p,\,\bq]$ be a unit quaternion with inverse $\bhfp^{*} = [p,\,-\bq]$: 
this requires $\bhfp \cast\bhfp^{*} = [p^2 + q^2,\,0] = [1,0]$ for which we need 
$p^2 + q^2 = 1$. For a pure quaternion $\bfr = [0,\,\br]$ there exists a 
transformation from $\bfr = [0,\,\br] \to \bfR = [0,\,\bR]$
\bel{r1}
\bfR = \bhfp\cast\bfr\cast\bhfp^{*}\,.
\ee
This associative product can explicitly be written as
\bel{r2}
\bfR = \bhfp\cast\bfr\cast\bhfp^{*} 
= [0,\, (p^{2}-q^{2})\br + 2p(\bq\times\br)+ 2\bq(\br\cdot\bq)]\,.
\ee
Choosing $p=\pm\cos\shalf\theta$ and $\bq = \pm\,\bhn\sin\shalf\theta$, where $\bhn$ 
is the unit normal to $\br$, we find that 
\bel{r3}
\bfR = \bhfp\cast\bfr\cast\bhfp^{*} = [0,\, \br\cos\theta + (\bhn\times\br) \sin\theta] 
\equiv O(\theta\,,\bhn)\br\,,
\ee
where
\bel{r4}
\bhfp = \pm [\cos\shalf\theta,\,\bhn\sin\shalf\theta]\,.
\ee
Equation (\ref{r3}) is the famous \textit{Euler-Rodrigues formula} for the rotation 
$O(\theta\,,\bhn)$ by an angle $\theta$ of the 3-vector $\br$ about its normal $\bhn$\,; 
the quantities $\theta\,,\bhn$ are called the Euler parameters. The elements of the unit 
quaternion $\bhfp$ are the Cayley-Klein parameters\footnote{The Cayley-Klein 
parameters of the quaternion $\bfq = [\alpha,\,\bchi]$ of \S\ref{lagev} are given by
$$
\bhfq = \left[\frac{\alpha}{\alpha^2 + \chi^2},\, \frac{\bchi}{\alpha^2 + \chi^2}\right]\,.
$$} 
which are related to the Euler angles \cite{Whitt44}.
\begin{lemma}\label{lemSU2}
The unit quaternions form a representation of the Lie group $SU(2)$.
\end{lemma}
\textbf{Proof\,:} From (\ref{quatdef1}), the matrix representation of a unit quaternion is
\bel{rot5-1}
{\sf J} =  p\,{\sf I} - i \bq\cdot\bsigma = 
\left(\begin{array}{cc}
~p - i q_3 & - i q_1- q_2 \\
- i q_1+ q_2 & p + i q_3
\end{array}\right)
\ee
where ${\sf J}\in SU(2)$\,; that is, ${\sf J}$ is a unitary $2\times2$ matrix with unit 
determinant. Hence, we may rewrite the map (\ref{r1}) for quaternionic conjugation 
equivalently in terms of the Hermitian Pauli spin matrices as 
\bel{rot6}
\bR\cdot\bsigma = {\sf J}\,\br\cdot\bsigma {\sf J}^{\dagger}
\ee
This is the standard representation of $SO(3)$ rotations as a double covering $(\pm {\sf J})$ 
by  $SU(2)$ matrices, which is now seen to be equivalent to quaternionic multiplication\,: 
for more discussion of this theorem see a modern treatise on mechanics such as \cite{MaRa1994}. 
The $(\pm)$ in the Cayley-Klein parameters reflects the 2:1 covering of the map $SU(2)\mapsto 
SO(3)$. \hspace{5cm}$\blacksquare$
\par\medskip
To investigate the  map (\ref{r1}) when $\bhfp$ is time-dependent, the Euler-Rodrigues formula 
in (\ref{r3}) can be written as 
\bel{rot2}
\bfR(t) = \bhfp\cast\bfr\cast\bhfp^{*}
\hspace{1cm}\Rightarrow\hspace{1cm}
\bfr = \bhfp^{*}\cast\bfR(t)\cast\bhfp\,.
\ee
Thus $\dot{\bfR}$ has a time derivative given by
\beq\label{rot3}
\dot{\bfR}(t) &=& \dot{\bhfp}\cast(\bhfp^{*}\cast\bfR\cast\bhfp)\cast\bhfp^{*}
+ \bhfp\cast(\bhfp^{*}\cast\bfR\cast\bhfp)\cast\dot{\bhfp}^{*}\nonumber\\
&=& \dot{\bhfp}\cast\bhfp^{*}\cast\bfR + \bfR\cast\bhfp\cast\dot{\bhfp}^{*}\nonumber\\
&=& (\dot{\bhfp}\cast\bhfp^{*})\cast\bfR + \bfR\cast(\dot{\bhfp}\cast\bhfp^{*})^{*}\nonumber\\
&=& (\dot{\bhfp}\cast\bhfp^{*})\cast\bfR - ((\dot{\bhfp}\cast\bhfp^{*})\cast\bfR)^{*}\,,
\eeq
having used the fact on the last line that because $\bfR$ is a pure quaternion, $\bfR^{*} 
= -\bfR$.
Because $\bhfp = [p,\,\bq]$ is of unit length, and thus $p\dot{p} + q\dot{q} = 0$, 
this means that $\dot{\bhfp}\cast\bhfp^{*}$ is also a pure quaternion 
\bel{rot4}
\dot{\bhfp}\cast\bhfp^{*} = [0,\,\shalf\bcapom_{0}(t)]\,.
\ee
The 3-vector entry in (\ref{rot4}) defines the angular frequency $\bcapom_{0}(t)$ as 
$\bcapom_{0} = 2(-\dot{p}\bq +\dot{\bq}p - 
\dot{\bq}\times\bq)$ thereby giving the well-known formula for the rotation of a rigid body
\bel{rot5}
\dot{\bR} = \bcapom_{0}\times\bR\,.
\ee 
For a Lagrangian particle, the equivalent of $\bcapom_{0}$ is the Darboux vector $\bD_{a}$ 
in Theorem \ref{abthm} of \S\ref{lagev}.

\section{\large Lagrangian evolution equations and the quaternion picture}\label{lagev}

\subsection{An ortho-normal frame and particle trajectories}\label{ortho-normal}

Having set the scene in \S\ref{rigid} by describing some of the essential properties of 
quaternions, it is now time to apply them to the Lagrangian relation (\ref{w-dyn}) 
between the two vectors $\bw$ and $\ba$ which we shall repeat here
\bel{wequn}
\frac{D\bw}{Dt} = \ba\,.
\ee
Through the  multiplication rule in (\ref{quatdef2}) quaternions appear in the 
decomposition of the 3-vector $\ba$ into parts parallel and perpendicular to another 
vector, which we choose to be $\bw$. This decomposition is expressed as
\bel{decom1}
\ba = \alpha_{a}\bw + \bchi_{a}\times\bw = [\alpha_{a},\,\bchi_{a}]\cast[0,\,\bw]\,,
\ee
where the scalar $\alpha_{a}$ and 3-vector $\bchi_{a}$ are defined as
\bel{la2a}
\alpha_{a} = w^{-1}(\bhw\cdot\ba)\,,\hspace{2cm}\bchi_{a} = w^{-1}(\bhw\times\ba)\,.
\ee
Equation (\ref{decom1}) thus shows that the quaternionic product is summoned in a natural 
manner. It is now easily seen that $\alpha_{a}$ is the growth rate of the scalar magnitude 
($w =|\bw|$) which obeys
\bel{la3}
\frac{Dw}{Dt} = \alpha_{a}w\,,
\ee
while $\bchi_{a}$, the swing rate of the unit tangent vector $\bhw = \bw w^{-1}$, satisfies 
\bel{la5}
\frac{D\bhw}{Dt} = \bchi_{a}\times \bhw\,.
\ee

\par\vspace{-6mm}\noindent
\bc
\begin{minipage}[c]{.75\textwidth}
\begin{pspicture}
\psframe(0,0)(5,5)
\thicklines
\qbezier(0,1)(4,2.5)(0,4)
\thinlines
\put(2.01,2.43){\makebox(0,0)[b]{$\bullet$}}
\put(1.4,2.5){\makebox(0,0)[b]{\scriptsize$(\bx_{1},t_{1})$}}
\thinlines
\put(2,2.5){\vector(0,1){1}}
\put(2,3.7){\makebox(0,0)[b]{$\bhw$}}
\put(2,2.5){\vector(-2,-1){1}}
\put(.7,1.8){\makebox(0,0)[b]{$\bhchi_{a}$}}
\put(2,2.5){\vector(1,0){1}}
\put(3.8,2.4){\makebox(0,0)[b]{$\bhw\times\bhchi_{a}$}}
\thicklines
\qbezier(7,1)(6,2.5)(8,4)
\thinlines
\put(6.77,2.45){\makebox(0,0)[b]{$\bullet$}}
\put(6.2,2.5){\makebox(0,0)[b]{\scriptsize$(\bx_{2},t_{2})$}}
\thicklines
\qbezier[50](2,2.5)(3,.5)(6.7,2.5)
\thinlines
\put(6.73,2.5){\vector(1,4){.3}}
\put(7,3.8){\makebox(0,0)[b]{$\bhw$}}
\put(6.7,2.5){\vector(4,-1){1}}
\put(8.4,2.8){\makebox(0,0)[b]{$\bhw\times\bhchi_{a}$}}
\put(6.7,2.5){\vector(4,1){1}}
\put(8.1,2.1){\makebox(0,0)[b]{$\bhchi_{a}$}}
\put(3,1.2){\vector(1,0){.6}}
\put(4,.7){\makebox(0,0)[b]{\small tracer particle trajectory}}
\put(4.5,1.4){\vector(4,1){.6}}
\thinlines
\end{pspicture}
\end{minipage}
\ec
\bc
\vspace{-1mm}
\begin{minipage}[r]{\textwidth}
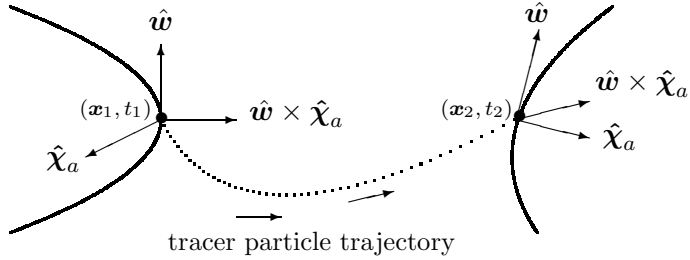
\textbf{Figure 1:} {\small The dotted line represents the tracer particle $(\bullet)$ path  
moving from $(\bx_{1},t_{1})$ to $(\bx_{2},t_{2})$. The solid curves represent lines of constant 
$\bw$ to which $\bhw$ is a unit tangent vector. The  orientation of the quaternion-frame
$(\bhw,\,\bhchi_{a},~\bhw\times\bhchi_{a})$ is  shown at the two space-time points; note that 
this is not the Frenet-frame corresponding to the particle path but to lines of constant $\bw$.}
\end{minipage}
\ec
\par\medskip\noindent
Now define the two quaternions
\bel{ls6}
\bfq_{a} 
= [\alpha_{a},\,\bchi_{a}]\,,
\hspace{2cm}
\bfw = [0,\,\bw]\,,
\ee
where $\bfw$ is a pure quaternion.  Then (\ref{wequn}) can automatically be re-written 
equivalently in the quaternion form
\bel{lem1}
\frac{D\bfw}{Dt} 
= \bfq_{a}\cast\bfw\,.
\ee
Moreover, if $\ba$ is differentiable in the Lagrangian sense (see (\ref{a-dyn})) 
\bel{aequn}
\frac{D\ba}{Dt} = \bdb\,,
\ee
then, exactly as for $\bfq_{a}$, a quaternion $\bfq_{b}$ can be defined which is based on the 
variables
\bel{alphachibdef}
\alpha_{b} = w^{-1}(\bhw\cdot\bdb)\,,
\hspace{3cm}
\bchi_{b} = w^{-1}(\bhw\times\bdb)\,,
\ee
where
\bel{qbdef}
\bfq_{b} = [\alpha_{b},\,\bchi_{b}]\,.
\ee
It is now clear that there exists a similar decomposition for $\bdb$ as that for $\ba$ as 
in (\ref{decom1}) 
\bel{la9}
\frac{D^{2}\bfw}{Dt^{2}} 
= [0,\,\bdb] 
= [0,\, \alpha_{b}\bw + \bchi_{b}\times\bw ] 
= \bfq_{b}\cast\bfw \,.
\ee
Using the associativity property, compatibility of (\ref{la9}) 
and (\ref{lem1}) implies that ($w = |\bw| \neq 0$)
\bel{la10}
\left(\frac{D\bfq_{a}}{Dt} + \bfq_{a}\cast\bfq_{a} 
-\bfq_{b}\right)\cast\bfw = 0\,,
\ee
which establishes a \textit{Riccati relation} between $\bfq_{a}$ and
$\bfq_{b}$
\bel{Ric1}
\frac{D\bfq_{a}}{Dt} + \bfq_{a}\cast\bfq_{a} = \bfq_{b}
\,,
\ee
whose components yield
\bel{Ric1-comp}
\frac{D}{Dt} [\alpha_{a},\,\bchi_{a}]
+ [\alpha_{a}^2-\chi_{a}^2,\,2\alpha_{a}\bchi_{a}] 
= [\alpha_{b}\,,\chi_{b}]
\,,
\ee
where $\chi_{a} = |\bchi_{a}|$. 
From (\ref{Ric1}), or equivalently (\ref{Ric1-comp}), there follows the first main result 
of the paper:
\begin{theorem}\label{abthm}
The ortho-normal quaternion-frame $(\bhw,\,\bhchi_{a},\,\bhw\times\bhchi_{a})\in SO(3)$ 
has Lagrangian time derivatives expressed as
\beq\label{abframe3}
\frac{D\bhw}{Dt}&=& \bD_{a}\times\bhw\,,\\
\frac{D(\bhw\times\bhchi_{a})}{Dt} &=& \bD_{a}\times(\bhw\times\bhchi_{a})\,,\label{abframe4}
\\
\frac{D\bhchi_{a}}{Dt} &=& \bD_{a}\times\bhchi_{a}\,,\label{abframe5}
\eeq
where the Darboux angular velocity vector $\bD_{a}$ is defined as
\bel{abframe6}
\bD_{a} = \bchi_{a} + \frac{c_{b}}{\chi_{a}}\bhw\,,\hspace{2cm}
c_{b} = \bhw\cdot(\bhchi_{a}\times\bchi_{b})\,.
\ee
\end{theorem}
\par\smallskip\noindent
\textbf{Remark\,:} The Darboux vector $\bD_{a}$ sits in a two-dimensional plane and is driven 
by the vector $\bdb$ which sits in $c_{b}$ in (\ref{abframe6}). The analogy with rigid body 
rotation expressed in (\ref{rot5}) is clear.
\par\smallskip\noindent
\textbf{Proof\,:} To find an expression for the Lagrangian time derivatives of the components of 
the frame  $(\bhw,\,\bhchi_{a},\,\bhw\times\bhchi_{a})$ requires the derivative of $\bhchi_{a}$. 
To find this it is necessary to use the fact that the 3-vector $\bdb$ can be expressed in this 
ortho-normal frame as the linear combination
\bel{b1}
w^{-1}\bdb = \alpha_{b}\,\bhw + c_{\,b}\bhchi_{a} + d_{\,b}(\bhw\times\bhchi_{a})\,.
\ee
where $c_{\,b}$ is defined in (\ref{abframe6}) and $d_{\,b} = -\,(\bhchi_{a}\cdot\bchi_{b})$.
The 3-vector product $\bchi_{b} = w^{-1}(\bhw\times\bdb)$ yields 
\bel{bchibdef}
\bchi_{b} = c_{\,b}\,(\bhw\times\bhchi_{a}) - d_{\,b}\bhchi_{a}\,.
\ee
When split into components, equation (\ref{Ric1-comp}) becomes
\bel{abframe0}
\frac{D\alpha_{a}}{Dt} = \chi_{a}^{2} - \alpha_{a}^{2} + \alpha_{b}\,
\ee
and
\bel{abframe1}
\frac{D\bchi_{a}}{Dt} = - 2\alpha_{a}\bchi_{a} + \bchi_{b}\,.
\ee
From the latter it is easily seen that
\bel{scalarchi}
\frac{D\chi_{a}}{Dt} = -2\alpha_{a}\chi_{a} - d_{b}\,
\ee
from which it follows
\bel{abframe2}
\frac{D\bhchi_{a}}{Dt} = c_{b}\chi_{a}^{-1}(\bhw\times\bhchi_{a})\,,
\hspace{2cm}
\frac{D(\bhw\times\bhchi_{a})}{Dt} = \chi_{a}\,\bhw - c_{b}\chi_{a}^{-1}\bhchi_{a}\,,
\ee
which gives equations (\ref{abframe3})-(\ref{abframe6}).\hspace{8cm}$\blacksquare$
\par\smallskip
Theorem \ref{abthm} is the main result of the paper and is the equivalent for a Lagrangian particle undergoing fluid motion of the well-known formula (\ref{rot5}) for a rigid body undergoing rotation about its center of mass.

\subsection{The evolution of the $\bdb$-field in equation (\ref{aequn})}\label{bfield}

The Lagrangian rate of change of acceleration $D\ba/Dt = \bdb$ is important for tracking passive 
tracer particles. However, the vector $\bdb$ cannot be calculated directly from Ertel's Theorem in (\ref{a-dyn}).
As shown below, the Lagrangian evolution of $\bfq_{b}$ appearing in the quaternionic Riccati 
relation (\ref{Ric1}) may be described without approximation in terms of three arbitrary scalars. 
\begin{theorem}\label{b-field}
The Lagrangian time derivative of the quaternion $\bfq_{b}$ in the Riccati relation (\ref{Ric1}) can be expressed as 
\bel{qbev1}
\frac{D\bfq_{b}}{Dt} = \bfq_{a}\cast\bfq_{b} + \bPi_{a,b}\,,
\ee
\bel{Pidef}
\bPi_{a,b} 
= \lambda_{1}\bfq_{b} 
+\lambda_{2}\bfq_{a} 
+ \lambda_{3}{\rm Id}\,,
\ee
where $\lambda_{1}(\bx,\,t)\,,~\lambda_{2}(\bx,\,t)\,,~\lambda_{3}(\bx,\,t)$ are 
arbitrary  scalar functions and ${\rm Id} = [1,\,0]$ is the identity for the quaternions.
\end{theorem}
\par\medskip\noindent
\textbf{Remark:} Without further constraints $\lambda_{1}(\bx,\,t),\,~\lambda_{2}(\bx,\,t)$ 
and $\lambda_{3}(\bx,\,t)$ would be arbitrary.
\par\medskip\noindent
\textbf{Proof\,:} 
To establish (\ref{qbev1}), we differentiate the orthogonality relation $\bchi_{b}\cdot\bhw = 0$ 
and use the Lagrangian derivative of $\bhw$
\bel{press2}
\frac{D\bchi_{b}}{Dt} = \bchi_{a}\times\bchi_{b} + \bs_{0}\,,
\hspace{1cm}
\hbox{where}
\hspace{1cm}
\bs_{0} = \mu\bchi_{a} +  \lambda\,\bchi_{b}\,.
\ee
$\bs_{0}$ lies in the plane perpendicular to $\bhw$ in which $\bchi_{a}$ and $\bchi_{b}$ 
also lie and $\mu =\mu(\bx,t)$ and $\lambda = \lambda(\bx,t)$ are arbitrary scalars. 
Explicitly differentiating $\bchi_{b} = w^{-1}(\bhw\times\bdb)$ gives
\bel{alph1}
w^{-1}\bhw\left(\bchi_{a}\cdot\bdb\right) + \bs_{0} = 
- \alpha_{a}\bchi_{b} - \alpha_{b}\bchi_{a} + w^{-1}\bhw\left(\bchi_{a}\cdot\bdb\right) + 
w^{-1}\left(\bhw\times\frac{D\bdb}{Dt}\right)\,,
\ee
which can easily be manipulated into
\bel{alph2}
\bhw\times\left\{\frac{D\bdb}{Dt} - \alpha_{b}\,\ba  - \alpha_{a}\,\bdb\right\} = w\,\bs_{0}\,.
\ee
This means that
\bel{alph3}
\frac{D\bdb}{Dt} = \alpha_{b}\ba + \alpha_{a}\bdb + \bs_{0} \times\bw + \varepsilon\bw\,,
\ee
where $\varepsilon = \varepsilon(\bx,t)$ is a third unknown scalar in addition to $\mu$ 
and $\lambda$ in (\ref{press2}). Thus the Lagrangian derivative of 
$\alpha_{b} = w^{-1}(\bhw\cdot\bdb)$ is
\bel{alph4}
\frac{D\alpha_{b}}{Dt} 
= \alpha_{a}\alpha_{b} + \bchi_{a}\cdot\bchi_{b} + \varepsilon\,.
\ee
Lagrangian differential relations have now been found for $\bchi_{b}$ and $\alpha_{b}$, but at 
the price of introducing the triplet of unknown coefficients $\mu,~\lambda$, and $\varepsilon$ 
which are re-defined as
\bel{tetrad1}
\lambda =\alpha_{a} + \lambda_{1}\,,\hspace{1cm}
\mu = \alpha_{b} + \lambda_{2}\,,\hspace{1cm}
\varepsilon = -2\bchi_{a}\cdot\bchi_{b} + \lambda_{2}\alpha_{a} + \lambda_{1}\alpha_{b}+ \lambda_{3}\,.
\ee
The new triplet has been subsumed into the tetrad defined in (\ref{Pidef}). Then (\ref{press2}) 
and (\ref{alph4}) can again be written in the quaternion form (\ref{qbev1}).\hspace{3cm}$\blacksquare$
\par\smallskip
In \S\ref{constit} we shall discuss a approach for determining  $\bfq_{b}$ by introducing an 
approximate constitutive relation for the pressure Hessian ${\sf P}$ and auxiliary equations for 
the constitutive parameters.

\subsection{Frame dynamics and the Frenet equations}\label{Frenet}

\bc
\begin{minipage}[c]{.45\textwidth}
\begin{pspicture}
\psframe(0,0)(5,5)
\thicklines
\qbezier(0,0)(4,2.5)(0,5)
\thinlines
\put(2,2.5){\vector(0,1){1}}
\put(2,3.8){\makebox(0,0)[b]{$\bhw$}}
\thicklines
\thinlines
\put(2,2.45){\vector(1,0){1}}
\put(4,2.4){\makebox(0,0)[b]{$\bhw\times\bhchi_{a} = \nh$}}
\put(2,2.45){\vector(3,2){1}}
\put(3.6,3.1){\makebox(0,0)[b]{$\bhchi_{a} = \bh$}}
\end{pspicture}
\end{minipage}
\par\vspace{5mm}\noindent
\begin{minipage}[r]{\textwidth}
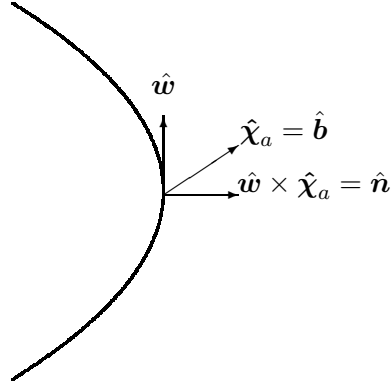
\textbf{Figure 2:} {\small The ortho-normal frame $(\bhw,~\bhw\times\bhchi_{a},~\bhchi_{a})$
as the Frenet-frame to lines of constant $\bw$.}
\end{minipage}
\ec
Modulo a rotation around the unit tangent vector $\bw$, with $\bhchi_{a}$ as the unit bi-normal 
$\bh$ and $\bhw\times\bhchi_{a}$ as the unit principal normal $\nh$, the matrix ${\sf F}$ can be 
formed
\bel{frmx}
{\sf F} = \left(\bhw^{T},\,(\bhw\times\bhchi_{a})^{T},\,\bhchi_{a}^{T}\right)\,,
\ee
and (\ref{abframe3})--(\ref{abframe5}) can be re-written as
\bel{frame3a}
\frac{D{\sf F}}{Dt} = {\sf A}{\sf F}\,,\hspace{2cm}
{\sf A} = \left(
\begin{array}{ccc}
0 &-\chi_{a} & 0\\
\chi & 0 &-c_{b}\chi_{a}^{-1}\\
0& c_{b}\chi_{a}^{-1} & 0
\end{array}\right)\,.
\ee
For a space curve parameterized by arc-length $s$, 
then the Frenet equations relating $d{\sf F}/ds$ to the curvature $\kappa$ and the torsion $\tau$ 
of the line of constant $\bw$ are
\bel{frame3b}
\frac{d{\sf F}}{ds} = {\sf N}{\sf F}
\hspace{1.5cm}\hbox{where}
\hspace{1.5cm}
{\sf N} = \left(
\begin{array}{rrr}
0    & \kappa     & 0\\
-\kappa & 0         & \tau\\
0    & -\tau & 0
\end{array}
\right)\,.
\ee
It is now possible to relate the $t$ and $s$ derivatives of ${\sf F}$ given in (\ref{frame3a}) 
and (\ref{frame3b}). At any time $t$ the integral curves of the vorticity vector field define a 
space-curve through each point $\bx$. The arc-length derivative $d/ds$ is defined by
\bel{integ1a}
\frac{d}{ds} = \bhw\cdot\nabla\,.
\ee
The evolution of the curvature $\kappa$ and torsion $\tau$ of a vortex line may be obtained 
from Ertel's theorem in (\ref{ertel2}), expressed as the commutation of operators
\bel{frame3d}
\Big[\frac{d}{ds},\,\frac{D}{Dt}\Big] = \alpha_{a}\frac{d~}{ds}\,.
\ee
Applying this to ${\sf F}$ and using the relations (\ref{frame3a}) and (\ref{frame3b}) 
establishes the following Theorem
\begin{theorem}\label{frenetthm}
The matrices ${\sf N}$ and ${\sf A}$ satisfy the Lax equation that relates the evolution 
of the curvature $\kappa$ and the torsion $\tau$ to $\alpha_{a},\,\chi_{a}$ and $c_{b}$ 
defined in equations (\ref{decom1}) and (\ref{abframe6})
\bel{com1}
\frac{D{\sf N}}{Dt} - \alpha_{a}{\sf N} = \frac{d{\sf A}}{ds} + [{\sf A},\,{\sf N}]\,.
\ee
\end{theorem}
Thus, if $\bchi_{a} = 0$ the curvature $\kappa$ is stationary.

\section{\large Three further examples}\label{examples}

The quaternionic formulation can be applied to other situations, such as the stretching  of fluid 
line-elements, incompressible and compressible motion of Euler fluids and ideal MHD \cite{MB}. 
\par\smallskip\noindent
\textbf{(i)  The Euler fluid equations for incompressible flow in a rotating frame:} The 
velocity form of Euler's equations for an incompressible fluid flow in a frame rotating at 
frequency $\bcapom$ is 
\bel{ex1}
\frac{D\bu}{Dt} =
\underbrace{(\bu\times2\bcapom)}_{Coriolis} -\nabla p\,,\quad\hbox{with}\quad
{\rm div}\,\bu = 0\,.
\ee
Taking the curl yields ($\bom = {\rm curl}\,\bu $)
\bel{ex2}
\frac{D\bqt}{Dt} = \bqt\cdot\nabla \bu \,,\qquad\hbox{with}\qquad
\bqt = \rho^{-1}(\bom + 2\bcapom)\,.
\ee
The triad of vectors $(\bu,\,\bw,\,\ba)$  in this case represents
$(\bu,\,\bqt,\,\bqt\cdot\nabla\bu)$ with  $\bom  = \hbox{curl}\,\bu$ and
$\hbox{div}\,\bu = 0$. Then Ertel's theorem becomes
\bel{ex3}
\frac{D}{Dt}(\bqt\cdot\nabla\bmu) = \bqt\cdot\nabla\left(\frac{D\bmu}{Dt}\right)\,.
\ee
Upon taking $\bmu=\bu$ in Ertel's theorem as in \cite{Ohk93} and using the motion equation 
gives a relation in the moving frame
\beq\label{ex5}
\frac{D}{Dt}(\bqt\cdot\nabla\bu)
&=& \bqt\cdot\nabla\big(\bu\times2\bcapom -\nabla p\big)\nonumber\\
&=&  -{\sf P}\bqt + \bqt\cdot\nabla\big(\bu\times2\bcapom\big)\,,
\eeq
where ${\sf P}$ is the Hessian matrix of the pressure defined by 
\bel{hess1}
{\sf P} = \frac{\partial^{2}p}{\partial x_{i}\partial x_{j}}\,.
\ee
Thus (\ref{ex3}) and (\ref{ex5}) identify $\ba$ and $\bdb$  as $\ba = \bqt\cdot\nabla\bu$ 
and $\bdb = -{\sf P}\bqt + \bqt\cdot\nabla\big(\bu\times2\bcapom\big)$; the particles are now 
fluid packets and not passive tracer particles. The divergence-free constraint
$\hbox{div}\,\bu = 0$ implies that 
\bel{ell1}
-\Delta p = u_{i,j}u_{j,i} 
- {\rm div}\,\big(\bu\times2\bcapom\big)
=  {\rm tr}\,{\sf S}^{2} - \shalf \omega^{2} - {\rm div}\,\big(\bu\times2\bcapom\big)\,.
\ee
Equation (\ref{ell1}) places an implicit condition upon the relation between the strain matrix 
${\sf S}$ and the pressure Hessian ${\sf P}$ in addition to the Riccati equation (\ref{la10}). 
This situation has been discussed at greater length in \cite{GHKR} in the absence of rotation.
\par\bigskip\noindent
\textbf{(ii) Euler's equations for a barotropic compressible fluid:} The pressure of a barotropic 
compressible fluid is a function of its mass density $\rho$, so it satisfies $\nabla\rho\times\nabla
p=0$. The velocity form of Euler's equations for incompressible fluid motion in a frame rotating 
at frequency $\bcapom$ is 
\bel{ex6}
\frac{D\bu}{Dt} = -\,\frac{1}{\rho}\nabla p(\rho) =: -\,\nabla h(\rho)\,,
\ee
\bel{rhoequn}
\frac{D\rho}{Dt} + \rho\,{\rm div}\,\bu = 0\,.
\ee
Taking the curl yields
\bel{ex7}
\frac{D\bqr}{Dt} = \bqr\cdot\nabla\bu \,,\quad\hbox{with}\quad
\bqr = \bom/\rho\quad\hbox{and}\quad\bom = {\rm curl}\,\bu \,.
\ee
Then Ertel's theorem takes the same form as above, and the second Lagrangian time derivative yields
the Ohkitani relation for a barotropic compressible fluid,
\bel{ex8}
\frac{D^2\bqr}{Dt^2}
= \frac{D}{Dt}(\bqr\cdot\nabla \bu)
= -\,\bqr\cdot\nabla\left(\nabla h(\rho)\right) \equiv \bdb\,,
\ee
in terms of the Hessian of its specific enthaply, $h(\rho)$. This has the same form 
as for incompressible fluids, except the acceleration term $\bdb= -\,\bqr\cdot\nabla
\left(\nabla h(\rho)\right)$ has its own dynamical equation. Thus, the methods in 
\cite{GHKR,Gibbon02} also apply for barotropic fluids. For isentropic compressible 
fluids, the situation is more complicated. 
\par\smallskip\noindent
\textbf{(iii) The equations of incompressible ideal MHD:} These are
\bel{F1}
\frac{D\bu}{Dt} = \bB\cdot\nabla\bB - \nabla p\,,
\ee
\bel{F2}
\frac{D\bB}{Dt} = \bB\cdot\nabla\bu\,,
\ee
together with $\mbox{div}\,\bu = 0$ and $\mbox{div}\,\bB = 0$. The pressure
$p$ in (\ref{F1}) is $p = p_{f} + \frac{1}{2}B^{2}$ where $p_{f}$ is the 
fluid pressure. Elsasser variables are defined by the combination
\bel{F3}
\bv^{\pm} = \bu \pm \bB\,.
\ee
The existence of two velocities $\bv^{\pm}$ means that there are two material 
derivatives
\bel{F3a}
\frac{D^{\pm}}{Dt} = \frac{\partial~}{\partial t} + \bv^{\pm}\cdot\nabla\,.
\ee
In terms of these, (\ref{F1}) and (\ref{F2}) can be rewritten as 
\bel{F4}
\frac{D^{\pm}\bv^{\mp}}{Dt} = - \nabla p\,,
\ee
with the magnetic field $\bB$ satisfying ($\mbox{div}\,\bv^{\pm} = 0$)
\bel{F5}
\frac{D^{\pm}\bB}{Dt} = \bB\cdot\nabla \bv^{\pm}\,.
\ee
Thus we have a pair of triads 
$(\bv^{\pm},\,\bB,\,\ba^{\pm})$ with $\ba^{\pm} = \bB\cdot\nabla \bv^{\pm}$, 
based on Moffatt's identification of the $\bB$-field as the important 
stretching element \cite{HKM1}. From \cite{GHKR,Gibbon02} we also have 
\bel{F8}
\frac{D^{\pm}\ba^{\mp}}{Dt} = - {\sf P}\bB\,,
\ee
where $\bdb^{\pm} = -{\sf P}\bB$. With two quartets $(\bv^{\pm},\,\bB,\,\ba^{\pm}\,,\bdb)$, 
the results of Section 2 follow, with two Lagrangian derivatives and two Riccati 
equations 
\bel{Ric2}
\frac{D^{\mp}\bfq_{a}^{\pm}}{Dt} + \bfq_{a}^{\pm}\cast\bfq_{a}^{\mp} = \bfq_{b}\,.
\ee
In consequence, MHD-quaternion-frame dynamics needs to be interpreted in terms of two sets 
of ortho-normal frames $\left(\bhB,\,\bhchi^{\pm},\,\bhB\times\bhchi^{\pm}\right)$ acted 
on by their opposite Lagrangian time derivatives. 
\beq\label{frameMHD1}
\frac{D^{\mp}\bhB}{Dt}&=& \bD^{\mp}\times\bhB\,,
\\
\frac{D^{\mp}}{Dt}(\bhB\times\bhchi^{\pm}) &=& \bD^{\mp}\times(\bhB\times\bhchi^{\pm})\,,
\\
\frac{D^{\mp}\bhchi^{\pm}}{Dt} &=& \bD^{\mp}\times\bhchi^{\pm}\,,
\eeq
where the pair of Elsasser Darboux vectors $\bD^{\mp}$ are defined as
\bel{frameMHD2}
\bD^{\mp} = \bchi^{\mp} - \frac{c_{B}^{\mp}}{\chi^{\mp}}\bhB\,,
\hspace{2cm}
c_{B}^{\mp} = \bhB\cdot[\bhchi^{\pm}\times(\bchi_{pB} + \alpha^{\pm}\bchi^{\mp})]\,.
\ee

\section{\large Approximate constitutive relations for the pressure Hessian}\label{constit}

In this section, we discuss an approach for determining  $\bfq_{b}$ by introducing an approximate 
constitutive relation for the pressure Hessian ${\sf P}$ in the Euler equations and auxiliary 
equations for these constitutive parameters. Recall Euler's familiar equations for an 
incompressible fluid flow with velocity $\bu$, written as
\bel{Eul-vel}
\frac{D\bu}{Dt} =
-\nabla p\,,\quad\quad\hbox{with}\quad\quad{\rm div}\,\bu = 0\,.
\ee
Taking the gradient yields the matrix Riccati equation
\bel{pressure-velgrad-eqns}
\frac{D{\sf M}}{Dt} + {\sf P} + {\sf M}^2 = 0\,,
\ee
where the velocity gradient tensor ${\sf M}=\nabla\bu$ has Cartesian components 
$M_{ij}=\partial u_{j}/\partial x^{i} = u_{j,i}$ and the (symmetric) pressure Hessian 
$ P =\nabla\nabla{p}$ has Cartesian components ${\sf P}_{ij}=\p^2{p}/\p{x^i}\p{x^j}$.
\par\smallskip
Because of the incompressibility condition ${\rm div}\,\bu = 0$, the trace of the velocity 
gradient tensor
${\rm tr}\,{\sf M}$ vanishes, thereby requiring ${\rm tr}\,{\sf P} = - {\rm tr}\,({\sf M}^2)$, 
which is a Poisson equation for the pressure. For laminar flow in a bounded domain, the Poisson 
equation determines both the pressure in the exact Euler equations  and its Hessian appearing in 
the velocity gradient equations (\ref{pressure-velgrad-eqns}). 
\par\smallskip
However, in turbulent flows, modern diagnostics for both numerical simulations and fluid measurements
make extensive use of average values of the velocity gradients moving in a coarse-grained volume 
element following the mean flow \cite{ZeLaMcRoKoLa2003}. The process of averaging following a 
fluid parcel is called \emph{Lagrangian averaging}. By its definition, Lagrangian averaging 
commutes with the material derivative, but with not the spatial gradient. In contrast, Eulerian 
averaging does the opposite.  With its two Eulerian spatial gradients, the Lagrangian averaged 
Hessian is a challenging object to compute.  Several attempts have been made to model the 
Lagrangian averaged pressure Hessian in (\ref{pressure-velgrad-eqns}) by introducing a 
constitutive closure for it. This idea goes back to L\'eorat \cite{Le1975}, Vieillefosse 
\cite{Vi1984} and Cantwell \cite{Ca1992} who assumed that the Eulerian pressure Hessian ${\sf P}$ 
is isotropic\,: see also \cite{OoMaSoCh1995,MaOoChSo1998}. This assumption results in the 
\emph{restricted Euler equations} (\ref{Eul-vel}) and (\ref{pressure-velgrad-eqns}) with 
\bel{Hess-relA}
{\sf P} =  -\,\frac{{\sf Id}}{3}\,{\rm tr}\,({\sf M}^2)\,,
\ee
where ${\sf Id}_{ab}=\delta_{ab}$ and ${\rm tr}\,({\sf Id})=3$  in three dimensions, so that taking
the trace satisfies the relation required for incompressibility \cite{GrSp1995}. Conversely, one may
assume that the Lagrangian pressure Hessian is isotropic. The latter assumption underlies the mean 
flow features of the \emph{tetrad model} of Chertkov, Pumir and Shraiman \cite{ChPuSh1999}; see 
Chevillard and Meneveau \cite{ChMe2006} for a recent review of this approach and results on its
use in turbulence diagnostics.
\par\smallskip
A more general model that encompasses the mean flow features of both the restricted Euler equations
and the tetrad model emerges from the transformation properties of the pressure Hessian under the 
Lagrangian flow map. The pressure Hessian transforms from Eulerian to Lagrangian coordinates by the
flow map $\phi_t:\,X\to x(t)$ in which $x(t)$ denotes the present position of a certain fluid particle,
at time $t$, that started initially at position $X=x(0)$ at time $t=0$. This flow map preserves volume.
Consequently, it is invertible and its Jacobian, the deformation gradient tensor, 
${\sf D}^i_A(X,t) = \p x^i /\p X^A$ has unit determinant $\det({\sf D}) = 1$. If it were frozen 
into the flow, the pressure Hessian would transform under the flow map as
\bel{distance-rel}
\frac{\p^2 p(t)}{\p x^i\p x^j}\,dx^i(t)\otimes dx^j(t) 
= \phi_t\circ\bigg(
\frac{\p^2 p(0)}{\p X^A\p X^B}\,dX^A\otimes dX^B\bigg)\,.
\ee
This is the way that a Riemannian metric ${\sf G}$ transforms under a time-dependent change of 
spatial coordinates. Namely, the transformation of a Riemannian metric gives the evolving 
quantity  ${\sf G}(t)$ in terms of the reference metric ${\sf G}(0)$ and the change of basis 
governed by the evolution of Jacobian matrix  $({\sf D}^{-1})^A_i(t)=\p X^A /\p x^i$ of the 
\emph{inverse} flow map in tensor index notation as
\bel{Hess-relB}
{\sf G}_{ij}(t) = {\sf G}_{AB}(0)\,({\sf D}^{-1})^A_i(t)({\sf D}^{-1})^B_j(t)\,.
\ee
In short, a Riemannian metric in the Lagrangian reference configuration transforms under the 
flow map $\phi_t$ as
\bel{Hess-rel-short}
{\sf G}(t) = {\sf D}^{-1}(t){\sf G}(0){\sf D}^{-1}(t)\,,
\ee
where the material time derivative of ${\sf D}^{-1}(t)dx(t)=dx(0)$ determines the evolution 
of ${\sf D}^{-1}(t)$\footnote{The metric ${\sf G}(t)$ is called the Finger tensor in nonlinear 
elasticity. For its history and a modern application of the Finger tensor, see 
\cite{MaHu1983,YaMaOr2006,Tad03}.}  For the pressure Hessian to transform as a Riemannian 
metric and also to satisfy the Poisson equation for its trace, it must take the following 
algebraic form 
\bel{Hess-relC}
{\sf P} =  -\,\frac{{\sf G}}{{\rm tr}\,{\sf G}}{\rm tr}\,({\sf M}^2)\,.
\ee
Setting ${\sf G}(t)={\sf Id}$ recovers the restricted Euler equations of \cite{Vi1984,Ca1992}, while
setting ${\sf G}(0)={\sf Id}$ reformulates the mean flow part of the tetrad model \cite{ChPuSh1999}.
\par\smallskip
In fact, the Poisson equation for pressure is satisfied for \emph{any} choice of the nonsingular 
symmetric matrix ${\sf G}={\sf G}^T$ in equation (\ref{Hess-relC}) for the Hessian. Moreover, it 
may even be satisfied by choosing a linear combination of symmetric matrices in the form
\bel{Hess-rel-combo}
{\sf P} =  -\,\bigg[
\sum_{\beta=1}^N c_\beta
\frac{{\sf G}_\beta}{{\rm tr}\,{\sf G}_\beta}
\bigg]{\rm tr}\,({\sf M}^2)
\,,\quad\quad\hbox{with}\quad\quad
\sum_{\beta=1}^N c_\beta = 1\,,
\ee
so long as an evolutionary flow law is provided for each of the symmetric tensors ${\sf G}_\beta={\sf
G}_\beta^T$ with $\beta=1,\,\dots\,,N$. Any choice of these flow laws would also determine the 
evolution of the driving term $\bfq_{b}$ in the Riccati equation (\ref{Ric1}).

\section{\large Conclusions}\label{concl}

The review of rigid body rotations in \S\ref{rigid} shows that, if handled properly, quaternions 
have a computational advantage over Euler angle formulations. For example, the Euler-Rodrigues 
formula (\ref{r3}) arises from a simple multiplication of quaternions. When this approach is applied
to Lagrangian evolution equations, as in \S\ref{lagev}, it demonstrates that quaternions are a natural
and efficient way of calculating the orientation and angular velocity of Lagrangian particles in motion
through the introduction the concept of ortho-normal quaternion-frames travelling with each particle.
For any problem in this class, knowledge of the quartet of 3-vectors $(\bu,\,\bw,\,\ba,\,\bdb)$ is 
sufficient for the application of Theorem \ref{abthm}, which is the paper's main result. The complexity
of the various versions of the $3D$ Euler equations comes through the ortho-normal dynamics via the
pressure field which is itself coupled back through $\Delta p = -u_{i,j}u_{j,i}$. This has been discussed
more fully in \cite{AsKeKeGi1987,NaPu2005}. Adaptations of these ideas when more physics is added to
the Euler equations has been discussed in section  \S\ref{examples}. For the $3D$ Euler equations the
pressure affects the coefficient  $c_{b}$ in the Darboux vector $\bD_{ab}$ of Theorem \ref{abthm} through
the pressure Hessian ${\sf P}$. The pressure Hessian also figures prominently in Lagrangian averaged
models of velocity gradient dynamics in turbulence. These models employ approximate constitutive relations
for the pressure Hessian. The potential applicability of the present quaternionic frame description
in these Lagrangian averaged turbulence models was demonstrated in \S \ref{constit}.
\par\smallskip
It is also possible that the general formulation could be modified to include viscous effects, 
particularly if experimental data becomes available in a quaternionic format: 
the reader is referred to the review \cite{Falk01} and also to \cite{GKB}. While the current 
formulation depends only on $\bhom$ and not $\nabla\bhom$, the latter dependence could not be 
avoided if viscosity were included. An equivalent formulation for the compressible Euler 
equations (\cite{jts2,jts3}) may give a clue to the nature of the incompressible limit.

\par\vspace{2mm}\noindent
\textbf{Acknowledgements:} We thank Trevor Stuart, Christos Vassilicos \& Arkady Tsinober of 
Imperial College London. The work of DDH was partially supported by a Royal Society Wolfson 
award and by the US Department of Energy Office of Science ASCR program in Applied Mathematical 
Research. 

\par\vspace{-.25cm}

\bibliographystyle{unsrt}


\end{document}